\documentclass{article}

     \usepackage[preprint]{neurips_2025}

\usepackage[utf8]{inputenc} 
\usepackage[T1]{fontenc}    
\usepackage{hyperref}       
\usepackage{url}            
\usepackage{booktabs}       
\usepackage{amsfonts}       
\usepackage{nicefrac}       
\usepackage{microtype}      
\usepackage{amsmath}
\usepackage{algorithm}
\usepackage{algpseudocode}
\usepackage{array}
\usepackage[capitalise]{cleveref}
\usepackage{graphicx}
\usepackage{multirow}
\usepackage{enumitem}
\usepackage{wrapfig}
\usepackage{colortbl}
\usepackage{tcolorbox}
\title{\emph{SepPrune}: Structured Pruning for Efficient Deep Speech Separation}

\author{
\begin{tabular}{c}
Yuqi Li\textsuperscript{1}\thanks{Equal Contribution.}, 
Kai Li\textsuperscript{2}\footnotemark[1],
Xin Yin\textsuperscript{3},
Zhifei Yang\textsuperscript{4}, 
Junhao Dong\textsuperscript{5} \\
Zeyu Dong\textsuperscript{1}, 
Chuanguang Yang\textsuperscript{1}, 
Yingli Tian\textsuperscript{6},
Yao Lu\textsuperscript{7}\thanks{Corresponding Author.}
\end{tabular} \\
\textsuperscript{1}Institute of Computing Technology, Chinese Academy of Sciences \\
\textsuperscript{2}Tsinghua University,
\textsuperscript{3}Zhejiang University,
\textsuperscript{4}Peking University \\
\textsuperscript{5}Nanyang Technological University,
\textsuperscript{6}The City University of New York,
\textsuperscript{7}A*STAR
}

\begin{document}

\maketitle

\begin{abstract}
Although deep learning has substantially advanced speech separation in recent years, most existing studies continue to prioritize separation quality while overlooking computational efficiency, an essential factor for low-latency speech processing in real-time applications. In this paper, we propose \emph{SepPrune}, the first structured pruning framework specifically designed to compress deep speech separation models and reduce their computational cost. \emph{SepPrune} begins by analyzing the computational structure of a given model to identify layers with the highest computational burden. It then introduces a differentiable masking strategy to enable gradient-driven channel selection. Based on the learned masks, \emph{SepPrune} prunes redundant channels and fine-tunes the remaining parameters to recover performance. Extensive experiments demonstrate that this learnable pruning paradigm yields substantial advantages for channel pruning in speech separation models, outperforming existing methods. Notably, a model pruned with \emph{SepPrune} can recover 85\% of the performance of a pre-trained model (trained over hundreds of epochs) with only one epoch of fine-tuning, and achieves convergence 36$\times$ faster than training from scratch. Code is available at \url{https://github.com/itsnotacie/SepPrune}.  
\end{abstract}

\section{Introduction}
\label{sec:intro}
Speech separation, the task of isolating individual speakers from a mixed audio signal, has made remarkable progress in recent years~\cite{jiang2025dual,li2024spmamba,subakan2021attention,subakan2022real,xu2024tiger,yang2022tfpsnet,zeghidour2021wavesplit,zhang2021transmask}. However, most speech separation methods have focused primarily on improving model performance. While these "performance-first" approaches are effective in controlled, laboratory settings, they fall short in real-world applications that demand low latency and constrained resource consumption, such as real-time voice communication~\cite{maryn2017mobile}, embedded auditory systems~\cite{briere2008embedded,das2014you}, etc. 

To bridge this efficiency gap, recent efforts~\cite{li2022efficient,xu2024tiger} have attempted to develop lightweight models through manual architectural design. However, such handcrafted models suffer from two fundamental limitations. First, they depend heavily on expert-driven architectural modifications and require substantial domain-specific knowledge. Second, and more importantly, these manual modifications are typically tailored to specific architectures, limiting their generalizability to other models. In light of the dual dilemma faced by manually designing architectures, this paper explores an alternative, non-invasive optimization strategy: \textit{model pruning}. 

Although model pruning has been shown to be effective in compressing vision and language models \cite{frankle2018lottery,lu2024reassessing,ma2023llm}, striking a balance between inference speed, memory usage, and accuracy, to the best of our knowledge, no pruning algorithm currently exists for end-to-end speech separation models. Unlike traditional vision~\cite{he2016deep,liu2021swin,qian2024reasoning,simonyan2014deep} or language~\cite{achiam2023gpt, touvron2023llama} models, speech separation models typically consist of three heterogeneous components: an audio encoder, a deep separation network, and an audio decoder. The computational complexity across these components is highly imbalanced. Consequently, indiscriminate pruning may damage already lightweight layers, leading to a collapse in model performance.

To address these challenges, we propose \emph{SepPrune}, the first structured pruning framework specifically designed for speech separation models.  \emph{SepPrune} consists of three stages. First, it performs a computational structural analysis on existing speech separation models to identify the layers that contribute most significantly to the overall computation. Next, it introduces a differentiable pruning mechanism using Gumbel-Softmax and a modified Straight-Through Estimator to build a set of differentiable channel binary masks to learn which channels should be kept. Finally, \emph{SepPrune} keeps the more important channels while removing the less important channels based on the binary masks, and fine-tunes the pruned model to recover the performance. Experiments show that \emph{SepPrune} not only significantly reduces the number of parameters and FLOPs of models, but also outperforms the previous state-of-the-art channel pruning methods~\cite{gao2024bilevelpruning,lin2020hrank} on the three benchmark datasets of LRS2-2Mix~\cite{li2022efficient}, Libri2Mix~\cite{cosentino2020librimix}, and EchoSet~\cite{xu2024tiger}. More notably, the pruned model obtained by \emph{SepPrune} can recover $86\%+$ of the performance of the original model trained for $493$ epochs with only $1$ epoch of fine-tuning, and it converges $36$ times faster than training from scratch.


In summary, our main contributions are as follows:
\begin{itemize}
    \item We introduce \emph{SepPrune}, the first pruning framework tailored specifically for deep speech separation models. \emph{SepPrune} performs structural computational analysis on the target model to determine the layers with the highest computational cost. Furthermore, \emph{SepPrune} introduces binary differentiable channel masks to select an optimal substructure. Based on the obtained masks, \emph{SepPrune} performs channel pruning and fine-tunes the remaining weights to recover performance.
    \item Extensive experiments demonstrate that \emph{SepPrune} outperforms existing channel pruning methods on three benchmark datasets (Libri2Mix, LRS2‑2Mix and EchoSet) and various backbones (A‑FRCNN‑12, TDANet, SuDoRM‑RF1.0x). It significantly reduces the complexity of the model while only causing minimal performance loss. Further experiments demonstrate its fast convergence and practical speedup effect.
\end{itemize}

\section{Related Works}
\label{sec:rw}

\subsection{Model Pruning} Model pruning is a widely used technique to compress pre-trained models by eliminating redundant parts, which can be roughly divided into three categories: weight pruning~\cite{bai2022dual,chen2023rgp,hoang2023revisiting,liu2022unreasonable,sun2023simple,wang2023ntk,wang2021recent,yang2025wanda++}, channel pruning~\cite{guo2020dmcp,he2017channel,ling2024slimgpt,li2016pruning,liu2019metapruning,ma2023llm,zhuang2018discrimination} and layer pruning~\cite{chen2018shallowing,li2024sglp,lu2022understanding,lu2024generic,lu2024reassessing,tang2023sr,wu2023efficient}. Specifically, in weight pruning, the unimportant weights are set to zero to reduce the total number of parameters. Although this method can achieve an extremely high compression rate, it depends on specialized hardware~\cite{han2016eie,park2016faster} for real speed-ups, so its inference speed gains in real-world deployments are often modest. In contrast, both channel pruning and layer pruning can achieve inference acceleration on standard computing devices. However, layer pruning removes the whole layer at once, which can severely impair model performance. Consequently, this paper focuses on channel pruning. It targets the channel dimension of layers, removing less important channels to reduce model size, while striving to preserve the model’s structure and performance.


\subsection{Speech Separation} The purpose of speech separation is to separate a single speech signal from a speech mixture. These methods can be roughly divided into two categories: time domain and time-frequency domain. Time domain methods directly utilize the original audio signal to achieve separation. For example, Conv-TasNet~\cite{luo2019conv} employs a linear encoder to create speech waveform representations optimized for speaker separation, with a linear decoder converting them back. A temporal convolutional network with stacked 1D dilated convolutional blocks is used to recognize masks and effectively capture long-term dependencies. DPT-Net~\cite{chen2020dual} introduces direct context-awareness in speech sequence modeling through an improved transformer that integrates recurrent neural networks into the original transformer. In contrast, time-frequency domain methods need to first convert the audio signal into a spectrogram representation using Short-Time Fourier Transform (STFT) to achieve separation. For instance, TF-GridNet~\cite{wang2023tf} employs stacked multi-path blocks containing intraframe spectral, sub-band temporal, and full-band self-attention modules to jointly exploit local and global spectro-temporal information for separation. BSRNN~\cite{luo2023music} explictly splits the spectrogram of the mixture into subbands and perform interleaved band-level and sequence-level modeling. While significant advances have been achieved in speech separation performance, current research mainly focuses on laboratory benchmarks while overlooking critical deployment requirements in practical systems, particularly the need for low-latency processing and computationally efficient operation.

\subsection{Efficient Speech Separation}
In real-world applications, speech separation models need to not only pursue separation quality, but also consider the computational efficiency for real-time processing. To this end, TDANet~\cite{li2022efficient} proposes an efficient lightweight architecture using top-down attention, achieving competitive performance with lower computational costs. Li et al.~\cite{li2024subnetwork} introduce a dynamic neural network that trains a large model with dynamic depth and width during the training phase and selects a subnetwork from it with arbitrary depth and width during the inference phase. Recently, Tiger~\cite{xu2024tiger} utilizes prior knowledge to divide frequency bands and compresses fre-quency information. We employ a multi-scale selective attention module to extract contextual features, while introducing a full-frequency-frame attention module to capture both temporal and frequency contextual information. Although these efficient speech separation methods have achieved promising results, their reliance on novel lightweight architecture designs leaves the critical challenge of compressing existing high-parameter models largely unaddressed.

\section{Preliminary}
Speech separation aims to extract individual speech signals of different speakers from a mixture, which can be formulated as:
\begin{equation}
    x=\sum_{1}^{C}s_i+\epsilon.
\end{equation}
$s_i\in\mathcal{R}^{1\times T}$ and $x\in\mathcal{R}^{1\times T}$ denote the waveform of speaker $i$ and a multi-speaker audio signal with the length $T$, respectively.  $\epsilon\in\mathcal{R}^{1\times T}$ denotes the noise signal and $C$ denotes the number of speakers. 

For speech separation tasks, most current state-of-the-art models~\cite{hu2021speech,li2022efficient,tzinis2020sudo,xu2024tiger} use a three-stage modular design of “an audio encoder $\rightarrow$ a separation network $\rightarrow$ an audio decoder”. Specifically, the audio encoder converts the mixed audio signal into a mixture audio representation. Subsequently, the separation network utilizes a deep neural network to produce a set of speaker-specific masks. Each target speech representation is then obtained by element-wise multiplying the mixture audio representation with its corresponding mask. Finally, the target waveform is reconstructed using the target speech representation through an audio decoder.



\section{Method}
\label{sec:method}
\subsection{Compressing Speech Separation Models via Channel Pruning}
\label{sec:Compressing Speech Separation Models via Channel Pruning}
This paper aims to slim a pre-trained speech separation model by pruning its channels. Given a $L$-layer pre-trained model, channel pruning aims to find a set of binary masks \begin{equation}
\mathcal{M}_{L\times1}=\{m_1,m_2,\cdots,m_L\},
\end{equation}
where each mask $m_l \in \{0,1\}^{C_l}$ corresponds to the $l$-th layer and $C_l$ is its number of channels. The objective is to mark each channel for removal ($0$) or retention ($1$) so as to reduce model complexity while maintaining its performance. To obtain the mask, a common paradigm in prior work is to minimize the loss $\mathcal{L}$ after pruning, which can be formulated as:
\begin{equation}
\min_{\Theta,\mathcal{M}} \; \mathbb{E}_{(\mathbf{x})}\Bigl[\mathcal{L}\bigl(f(\mathbf{x},\Theta , \mathcal{M}\bigr)\Bigr],
\end{equation}
where $\Theta$ and $\mathbf{x}$ represent the pre-training weights and input dataset respectively. However, the direct joint optimization of discrete masks and continuous weights is neither computationally tractable nor easy to converge. To this end, we decouple the joint optimization by first optimizing the masks and then fine-tuning the weights.
\begin{equation}
\underbrace{\min_{\Theta}}_{\text{Weight Learning}} \quad  \underbrace{\min_{\mathcal{M}} \; \mathbb{E}_{(\mathbf{x})}\Bigl[\mathcal{L}\bigl(f(\mathbf{x},\Theta, \mathcal{M}\bigr)\Bigr]}_\text{Mask Learning}.
\label{eq:loss}
\end{equation}
Although this objective formulated by \cref{eq:loss} may seem simple to implement, it still has two challenges that make it difficult to work in practice:
\begin{itemize}
    \item Masks are discrete variables and difficult to optimize using gradient descent;
    \item The number of mask combinations explodes, making optimization extremely difficult.
\end{itemize}
To address this, we introduce \emph{SepPrune}, which makes the masks optimizable and decouples their selection from the subsequent weight fine-tuning.

\begin{figure*}[t]
	\centering
    \centering
    \includegraphics[width=0.99\textwidth]{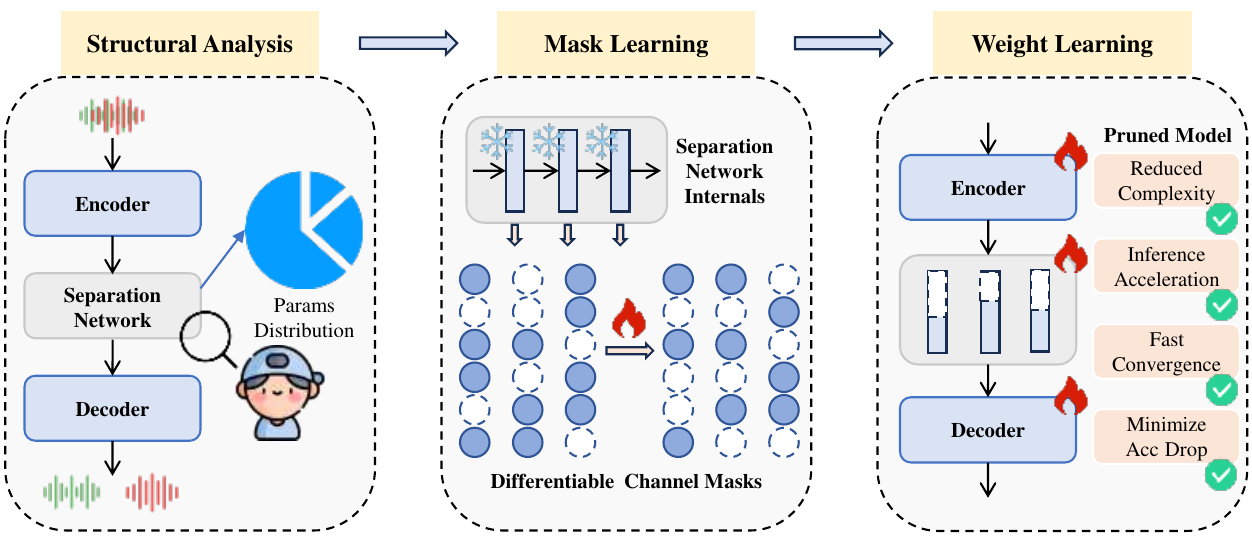} 
 \caption{The overall pipeline of the proposed \emph{SepPrune}.}
 \label{fig:pipeline}
\end{figure*}

\subsection{\emph{SepPrune}: Structured Channel Pruning via Differentiable Masks}
In this subsection, we delve into our \emph{SepPrune}. As illustrated in \cref{fig:pipeline}, our method consists of three core stages:
\begin{itemize}
    \item \textbf{Structural Computational Analysis}: Identify the layer contributing most to the overall computation.
    \item \textbf{Mask Learning with Gumbel-Softmax}: Learn a binary channel mask via Gumbel-Softmax to select a substructure that minimizes task loss.
    \item \textbf{Channel Pruning and Weight Refinement}: Perform channel pruning based on the obtained mask and fine-tune the remaining weights to recover performance.
\end{itemize}

\textbf{Structural Computational Analysis.}
Different from traditional convolutional neural networks~\cite{simonyan2014deep,he2016deep} and transformers\cite{liu2021swin,touvron2023llama,achiam2023gpt}, speech separation models usually consists of a set of an audio encoder, a deep separation network and an audio decoder. In speech separation models, the parameter distribution and computational complexity of different modules are often highly unbalanced. If we do not identify the “heavyweight” layers first and blindly prune all modules uniformly, we are likely to weaken the already lightweight layers or over-prune the key layers, resulting in a significant performance degradation. Therefore, we first perform structural computational analysis to these models. In this paper, we mainly use the TDANet~\cite{li2022efficient}, A-FRCNN~\cite{hu2021speech} and SudoRM-RF~\cite{tzinis2020sudo} to conduct experiments. Specifically, given a pre-trained model $f(\Theta)$, 
we use the widely-used protocols, i.e., number of parameters (denoted as Params) and required Float Points Operations (denoted as FLOPs), to evaluate model size and computational requirement. To ensure the reproducibility of the results, we uniformly utilize the ptflops\footnote{https://pypi.org/project/ptflops} to perform precise statistics on Params and FLOPs. As shown in \cref{tab:params}, we find that the separation network accounts for more than $82\%$ of the total parameters and $76\%$ of the FLOPs of these speech separation models. It can be seen that the separation network is the module with the greatest pruning benefit, so in this paper we mainly perform channel pruning on this module to minimize the computational overhead. 

\begin{table*}[t]
  \centering
  \caption{Statistics of the number of parameters and FLOPs of different models. SM denotes the separation network.}
  \resizebox{0.99\textwidth}{!}{
    \begin{tabular}{ccccccc}
    \toprule
    Model  & Total Params & Total FLOPs & Params of the SM & FLOPs of the SM & SM Parameter Ratio & SM FLOPs Ratio \\
    \midrule
    A-FRCNN-12 &  5.13 M & 28.58 GMac & 4.22 M & 26.56 GMac & 82.31\% & 92.94\% \\
    A-FRCNN-16 &  5.13 M & 37.44 GMac & 4.22 M & 35.42 GMac & 82.31\% & 94.59\% \\
    SuDoRM-RF1.0x & 2.72 M & 4.65 GMac & 2.35 M & 3.57 GMac & 86.50\% & 76.86\% \\
    TDANet & 2.35 M & 4.77 GMac & 2.29 M & 4.59 GMac & 97.44\% & 96.22\% \\
    TDANet Large & 2.35 M & 9.20 GMac & 2.29 M & 9.09 GMac & 97.44\% & 98.84\% \\
    \bottomrule
    \end{tabular}}
  \label{tab:params}%
\end{table*}%

\textbf{Mask Learning with Gumbel-Softmax.} After locating the parts with the most parameters, our next goal is to find the channels that need to be pruned (masked) by optimizing \cref{eq:loss}. As we mentioned in \cref{sec:Compressing Speech Separation Models via Channel Pruning}, the objective formulated by \cref{eq:loss} faces two critical challenges. First, the exhaustive search space for binary masks can be prohibitively large even at low pruning ratios. For instance, masking a layer with $128$ channels at $25\%$ sparsity requires evaluating $C_{128}^{32}$ possible solutions, making it difficult to use strategies such as evolutionary algorithms~\cite{lin2020channel} or reinforcement learning~\cite{wang2024rl}. It is worth noting that some layers of the speech separation model have $512$ or more channels, and more than one layer needs to be masked, which means that the search space is actually much larger than $C_{128}^{32}$. Besides, while gradient-based optimization would be ideal for this high-dimensional search space, the discrete nature of binary masks ($0/1$) fundamentally blocks gradient flow. To overcome both of these challenges, we introduce the Gumbel-Softmax~\cite{fang2024tinyfusion,fang2024maskllm,gumbel1954statistical,jang2016categorical} technique to convert discrete masks into differentiable “soft” probability distributions, allowing us to efficiently explore the exponential mask space using gradient descent. 

Specifically, let $F_{i}\in \mathcal{R}^{B\times C_{i} \times H_{i}}$ denote the feature representation of layer $i$, where $C_{i}$ is the number of channels, $B$ is the batch size and $H_{i}$ represents the feature length. To prune redundant channels, we assign a learnable importance score $\alpha_i \in \mathcal{R}^{C_i}$ for each layer, where each scalar $a_{i,j}$ represents the importance score of the $j$-th channel. Then we apply the Gumbel-Softmax technique to the weights $\alpha_i$:
\begin{equation}
\pi_{i}=\frac{\exp \left(\left(\log \left(\alpha_{i}\right)+g_{i}\right) / \tau\right)}{\sum_j \exp \left(\left(\log \left(\alpha_{j}\right)+g_{j}\right) / \tau\right)},
\end{equation}
where $g_{i} =\textnormal{-}\log(\textnormal{-}\log(\mathcal{U}))$ is the random noise drawn from the Gumbel distribution, with $\mathcal{U} \sim \text{Uniform}(0,1)$ and $\tau$ is a temperature term. Subsequently, we further utilize the improved Straight-Through Estimator~\cite{bengio2013estimating} to binarize $\pi_i$ to $m_i \in \{0,1\}^{C_i}$. In the backward propagation phase, we preserve the identity mapping of gradients while bounding their magnitudes within the interval [-1, 1] to mitigate the risk of the gradient explosion.
\begin{equation}
\begin{cases}
m_i = \frac{\operatorname{sign}(\pi_i-\epsilon)+1}{2}, & \text{forward propagation},\\[1ex]
\bigtriangledown_{\pi_i} = \operatorname{Clip}(\pi_i,-1,1) = \max\bigl(-1,\min(1,\pi_i)\bigr), & \text{backward propagation},
\end{cases}
\end{equation}
where $\epsilon$ is a hyperparameter used to control the masking (pruning) ratio. Leveraging the mask $m_i$ and the feature representation $F_{i}$, we can obtain the masked feature $\hat{F_{i}}=m_i \odot F_{i}$. For simplicity, we omit $\alpha_i$ and use $\mathcal{A}$ to represent the set of learned weight $\alpha_i$. Then $\mathcal{A}$ is derived by optimizing \cref{eq:mask} using gradient descent.
\begin{equation}
    \min_{\mathcal{A}} \; \mathbb{E}_{(\mathbf{x})}\Bigl[\mathcal{L}\bigl(f(\mathbf{x},\Theta, \mathcal{A}\bigr)\Bigr], \quad \mathcal{A} \leftarrow \mathcal{A} - \eta_\mathcal{A}\frac{\partial \mathcal{L}}{\partial \mathcal{A}}. 
    \label{eq:mask}
\end{equation}
Finally, we can obtain the $\mathcal{M}$ based on $\mathcal{A}$.

\textbf{Channel Pruning and Weight Refinement.} After obtaining the set of binary channel masks $\mathcal{M}$, we perform channel pruning by retaining the channels indexed $m_{i,j}=1$ and removing those with $m_{i,j}=0$. The pruned model is fine-tuned to recover performance degradation caused by channel removal. Let $\hat{\Theta}$ denote the parameters of the pruned model, initialized from the surviving weights of the original model. The optimization objective is:
\begin{equation}
\min_{\hat{\Theta}} \; \mathbb{E}_{(\mathbf{x})}\Bigl[\mathcal{L}\bigl(f(\mathbf{x},\hat{\Theta}\bigr)\Bigr], \quad \hat{\Theta} \leftarrow \hat{\Theta} - \eta_{\hat{\Theta}} \nabla_{\hat{\Theta}} \mathcal{L}.
\end{equation}
\section{Experiments}
\label{sec:exp}
\subsection{Model and Dataset}
We report the performance of \emph{SepPrune} on LRS2-2Mix~\cite{li2022efficient}, Libri2Mix~\cite{cosentino2020librimix} and EchoSet~\cite{xu2024tiger}. All of these datasets are at a sampling rate of $16$ kHz. We describe the datasets used in detail below. As for models, we utilize TDANet~\cite{li2022efficient}, A-FRCNN~\cite{hu2021speech} and SudoRM-RF~\cite{tzinis2020sudo} to conduct experiments. These models are widely-used in the speech separation community~\cite{li2024spmamba,xu2024tiger}.

\textbf{Libri2Mix.} Each mixture audio in Libri2Mix is built by randomly selecting from a subset of LibriSpeech’s train-$100$~\cite{panayotov2015librispeech} and mixing with uniformly sampled Loudness Units relative to Full Scale (LUFS)~\cite{series2011algorithms} between -$25$ dB and -$33$ dB. Each mix of sounds contains two different speakers and has a duration of $3$ seconds with a sample rate of $8$ kHz.

\textbf{LRS2-2Mix}\footnote{\url{https://drive.google.com/file/d/1dCWD5OIGcj43qTidmU18unoaqo_6QetW/view}} is created from the LRS2~\cite{afouras2018deep} corpus, with $20,000$ utterances in the training set, $5,000$ in validation, and $3,000$ in testing. Two audios of different speakers from varied scenes—each resampled to $16$ kHz—are randomly selected from the LRS2 corpus and mixed with signal‑to‑noise ratios sampled between –$5$ dB and $5$ dB. The data simulation follows the WSJ0‑2Mix protocol\footnote{\url{http://www.merl.com/demos/deep-clustering/create-speaker-mixtures.zip}}, and each mixture audio is 2 seconds. 
 
\textbf{EchoSet}\footnote{\url{https://huggingface.co/datasets/JusperLee/EchoSet}} is a speech separation dataset with various noise and realistic reverberation generated from SoundSpaces $2.0$~\cite{chen2022soundspaces} and Matterport3D~\cite{chang2017matterport3d}. It comprises $20,268$ training utterances, $4,604$ validation utterances, and $2,650$ test utterances. Each utterance lasts for 6 seconds. The two speakers’ utterances are overlaid with a random overlap ratio at a signal‑to‑distortion ratio (SDR) sampled between –$5$ dB and $5$ dB, and noises from the WHAM! corpus~\cite{wichern2019wham} are added. The noises are mixed with SDR sampled between -$10$ dB and $10$ dB.

\subsection{Training and Evaluation}
\label{sec:Training and Evaluation}
As for training the original models, to make a fair comparison with previous speech separation methods, we trained all models for 500 epochs in line with \cite{xu2024tiger}. It is worth noting that our pruning method does not actually require this step. Since there is no pre-trained model directly available, we train the model ourselves. The batch size is set to $1$ at the utterance level. We use the Adam optimizer~\cite{kingma2014adam} with an initial learning rate of $0.001$ and negative SI-SDR as the training loss~\cite{le2019sdr}. Besides, SDRi and SI-SDRi~\cite{vincent2006performance} are used for evaluation, with higher values indicating better performance. Once the best model has not been found for $15$ consecutive epochs, we adjust the learning rate to half of the previous one. In addition, if the best model has not been found for $30$ consecutive epochs, we stop the training early. As for mask learning, we set the initial learning rate to $0.1$ and train all masks for $500$ iterations. Since training the speech separation model is very expensive, in order to minimize the cost of mask learning, we only use $500$ iterations. We verify the effects of different iterations in \cref{sec:ablation}. Without specific instructions, we default to setting $\epsilon=0.7$. In addition, the hyperparameters used for fine-tuning the pruned model are consistent with the model training. The Params and FLOPs are calculated for one second of audio at $16$ kHZ. For all experiments, we used $8\times$NVIDIA V100 and $4\times$NVIDIA A100 for training and testing.

\subsection{Comparisons with State-of-The-Art Methods}
\label{sec:sota}
\begin{wrapfigure}{r}{9cm}
    \centering
    \vspace{-5mm}
    \includegraphics[width=0.6\textwidth]{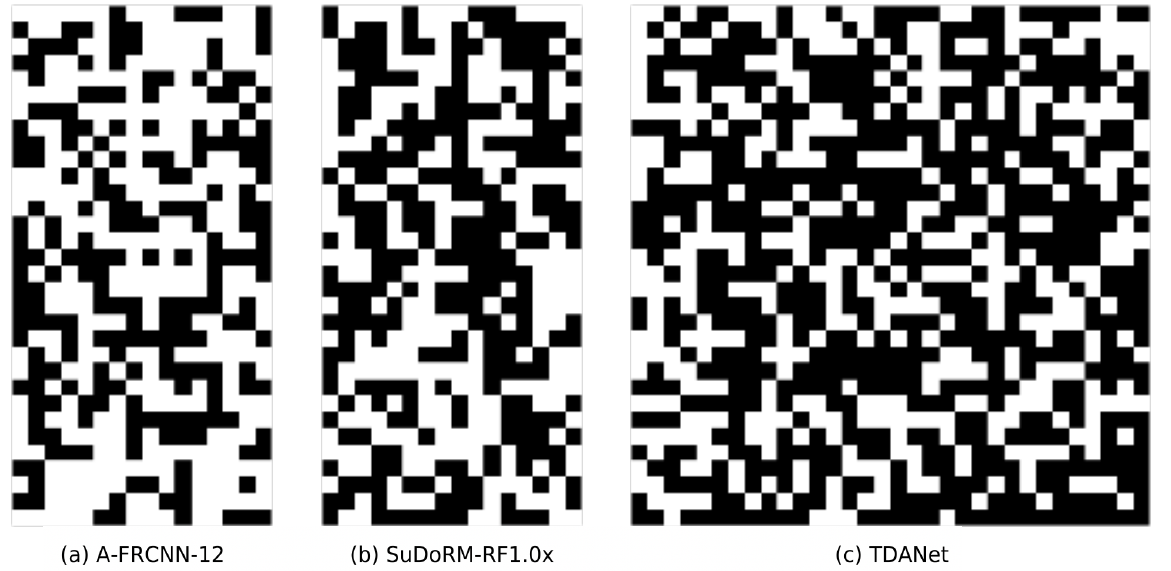} 
 \caption{Visualization of the obtained channel masks on the LRS2-2Mix dataset. For the convenience of visualization, we select the first layer on A-FRCNN-12, SuDoRM-RF1.0x and TDANet for visualization, and reshape the masks into $16\times32$, $16\times32$ and $32\times32$. }
 \label{fig:vis_mask} 
\end{wrapfigure}

To evaluate the effectiveness of \emph{SepPrune}, we compare our method with existing channel pruning methods (Random, Hrank~\cite{lin2020hrank} and UDSP~\cite{gao2024bilevelpruning}) on three benchmark datasets, including Libri2Mix, LRS2-2Mix, and EchoSet. Since these methods do not experiment on speech separation models, we reproduce them ourselves. As shown in Table~\ref{tab:performance}, we report the performance in terms of SDRi and SI-SDRi (in dB), along with the number of parameters (Params) and FLOPs after pruning. Across all datasets and model structures (TDANet~\cite{li2022efficient}, A-FRCNN-12~\cite{hu2021speech} and SuDoRM-RF1.0x~\cite{tzinis2020sudo}), \emph{SepPrune} consistently achieves superior performance under the same pruning ratio. For example, on LRS2-2Mix dataset with the A-FRCNN-12 model, \emph{SepPrune} not only outperforms all other pruning methods, but also achieves an SDRi of $12.59$ dB and an SI-SDRi of $12.25$ dB, both of which exceed the performance of the original model ($10.90$ dB and $10.50$ dB, respectively). Besides, on the most challenging EchoSet dataset, \emph{SepPrune} outperforms all baseline pruning strategies, yielding the highest SDRi and SI-SDRi in all cases. In addition, we visualize the learned masks in \cref{fig:vis_mask} to provide deeper insight into the selected channels. In summary, these results demonstrate the effectiveness and strong generalization ability of \emph{SepPrune}.

\begin{table}[t]
  \centering
  \caption{Performance comparison with existing pruning methods on Libri2Mix, LRS2-2Mix, and EchoSet. “-” indicates the original model. Bold denotes the best performance, and underline indicates the second-best. SDRi and SI-SDRi are recorded in dB.}
    \resizebox{0.99\textwidth}{!}{\begin{tabular}{c|c|cccc|cccc|cccc}
    \toprule
    \multirow{1.5}[4]{*}{Model} & \multirow{1.5}[4]{*}{Method} & \multicolumn{4}{c|}{LRS2-2Mix} & \multicolumn{4}{c|}{Libri2Mix} & \multicolumn{4}{c}{EchoSet} \\
\cmidrule{3-14}          &       & Params & FLOPs & SDRi  & SI-SDRi & Params & FLOPs & SDRi  & SI-SDRi & Params & FLOPs & SDRi  & SI-SDRi \\
    \midrule
    \multirow{5}[1]{*}{A-FRCNN-12} & -   & 5.13  & 28.58 & 10.90  & 10.50  & 5.13  & 28.58 & 15.54 & 15.03 & 5.13  & 28.58 & 8.56  & 7.66 \\
          & Random & 3.06  & 16.52 & 12.03 & 11.68 & 3.09  & 16.76 &  15.03     &   14.78    & 3.08  & 16.57 & 7.41  & 6.26 \\
          & Hrank & 3.06  & 16.52 & 12.45 & 12.11 & 3.09  & 16.76 &  \underline{15.32}     &   \underline{15.00}    & 3.08  & 16.57 & \underline{8.01}  & \underline{6.99} \\
          & UDSP  & 3.06  & 16.52 & \underline{12.49} & \underline{12.08} & 3.09  & 16.76 &   15.23    &    14.94   & 3.08  & 16.57 & 7.93  & 6.94 \\
          & \cellcolor[rgb]{ .867,  .922,  .969}\emph{SepPrune} & \cellcolor[rgb]{ .867,  .922,  .969}3.06  & \cellcolor[rgb]{ .867,  .922,  .969}16.52 & \cellcolor[rgb]{ .867,  .922,  .969}\textbf{12.59} &\cellcolor[rgb]{ .867,  .922,  .969}\textbf{12.25} & \cellcolor[rgb]{ .867,  .922,  .969}3.09  &\cellcolor[rgb]{ .867,  .922,  .969}16.76 &  \cellcolor[rgb]{ .867,  .922,  .969}\textbf{15.60}  &  \cellcolor[rgb]{ .867,  .922,  .969}\textbf{15.10} &\cellcolor[rgb]{ .867,  .922,  .969}3.08  & \cellcolor[rgb]{ .867,  .922,  .969}16.57 & \cellcolor[rgb]{ .867,  .922,  .969}\textbf{8.12}  & \cellcolor[rgb]{ .867,  .922,  .969}\textbf{7.13} \\
              \midrule
    \multirow{5}[1]{*}{TDANet} & -   & 2.35  & 4.77  & 12.74 & 12.45 & 2.35  & 4.77  & 13.42 & 12.92 & 2.35  & 4.77  & 8.78  & 7.93 \\
          & Random & 1.92  & 4.33  & 11.91 & 11.52 & 1.67  & 4.07  & 13.02 & 12.08 & 1.64  & 4.04  & 7.51  & 6.49 \\
          & Hrank & 1.92  & 4.33  & 12.33 & 12.01 & 1.67  & 4.07  & 13.23 & 12.46 & 1.64  & 4.04  & 7.69  & 6.67 \\
          & UDSP  & 1.92  & 4.33  & \underline{12.46} & \underline{12.26} & 1.67  & 4.07  & \underline{13.56} & \underline{12.79} & 1.64  & 4.04   & \underline{7.99}  & \underline{7.03}\\
          & \cellcolor[rgb]{ .867,  .922,  .969}\emph{SepPrune} & \cellcolor[rgb]{ .867,  .922,  .969}1.92  & \cellcolor[rgb]{ .867,  .922,  .969}4.33  & \cellcolor[rgb]{ .867,  .922,  .969}\textbf{12.72} & \cellcolor[rgb]{ .867,  .922,  .969}\textbf{12.41} & \cellcolor[rgb]{ .867,  .922,  .969}1.67  & \cellcolor[rgb]{ .867,  .922,  .969}4.07  & \cellcolor[rgb]{ .867,  .922,  .969}\textbf{13.70}  & \cellcolor[rgb]{ .867,  .922,  .969}\textbf{13.21} &\cellcolor[rgb]{ .867,  .922,  .969}1.64  & \cellcolor[rgb]{ .867,  .922,  .969}4.04  & \cellcolor[rgb]{ .867,  .922,  .969}\textbf{8.52}  &\cellcolor[rgb]{ .867,  .922,  .969}\textbf{7.62} \\
    \midrule
    \multirow{5}[2]{*}{SuDoRM-RF1.0x} & -   & 2.72  & 4.65  & 11.43 & 11.10  & 2.72  & 4.65  & 13.52 & 12.99 & 2.72  & 4.65  & 7.82  & 6.77 \\
          & Random & 1.54  & 2.91  & 9.53  & 9.06  & 0.98  & 2.09  & 12.14 & 11.82 & 1.11  & 2.28  & 6.52  & 5.74 \\
          & Hrank & 1.54  & 2.91  & \underline{10.29} & \underline{9.89}   & 0.98  & 2.09  & \underline{12.36} & 12.01 & 1.11  & 2.28  & 6.85  & 5.89 \\
          & UDSP  & 1.54  & 2.91  & 10.12 & 9.67 & 0.98  & 2.09  & 12.31 & \underline{12.04} & 1.11  & 2.28  & \underline{7.04}  & \underline{5.98} \\
          & \cellcolor[rgb]{ .867,  .922,  .969}\emph{SepPrune} & \cellcolor[rgb]{ .867,  .922,  .969}1.54  & \cellcolor[rgb]{ .867,  .922,  .969}2.91  & \cellcolor[rgb]{ .867,  .922,  .969}\textbf{10.37} & \cellcolor[rgb]{ .867,  .922,  .969}\textbf{9.98}  & \cellcolor[rgb]{ .867,  .922,  .969}0.98  &\cellcolor[rgb]{ .867,  .922,  .969}2.09  & \cellcolor[rgb]{ .867,  .922,  .969}\textbf{12.79} &\cellcolor[rgb]{ .867,  .922,  .969}\textbf{12.24} & \cellcolor[rgb]{ .867,  .922,  .969}1.11  &\cellcolor[rgb]{ .867,  .922,  .969}2.28  &\cellcolor[rgb]{ .867,  .922,  .969}\textbf{7.29}  & \cellcolor[rgb]{ .867,  .922,  .969}\textbf{6.07} \\
    \bottomrule
    \end{tabular}}
  \label{tab:performance}%
  \vspace{-3mm}
\end{table}%

\begin{table}[t]
  \centering
  \caption{Efficiency comparisons of the original model and pruned model. Experiments are conducted on the LRS2-2Mix dataset.}
  \resizebox{0.99\textwidth}{!}{
    \begin{tabular}{c|c|cccc|cccc}
    \toprule
    \multirow{1.5}[4]{*}{Model} & \multirow{1.5}[4]{*}{Type} & \multicolumn{4}{c|}{Train}    & \multicolumn{4}{c}{Inference } \\
\cmidrule{3-10}          &       & GPU Time  & GPU Memory  & Speed Up & Memory Savings & GPU Time  & GPU Memory  & Speed Up & Memory Savings \\
    \midrule
    \multirow{2}[1]{*}{A-FRCNN-12} & Original Model & 160.67ms & 3714MB & \multirow{2}[1]{*}{1.04$\times$} & \multirow{2}[1]{*}{38.50\%} & 104.23ms & 642MB & \multirow{2}[1]{*}{1.09$\times$} & \multirow{2}[1]{*}{2.20\%} \\
          & Pruned Model & 155.14ms & 2284MB &       &       & 95.38ms & 628MB &       &  \\
    \multirow{2}[0]{*}{TDANet} & Original Model & 435.75ms & 5130MB & \multirow{2}[0]{*}{1.06$\times$} & \multirow{2}[0]{*}{22.14\%} & 223.59ms & 640MB & \multirow{2}[0]{*}{1.08$\times$} & \multirow{2}[0]{*}{0.01\%} \\
          & Pruned Model & 409.56ms & 3994MB &       &       & 207.52ms & 639MB &       &  \\
    \multirow{2}[1]{*}{SuDoRM-RF1.0x} & Original Model & 157.43ms & 4084MB & \multirow{2}[1]{*}{1.04$\times$} & \multirow{2}[1]{*}{50.20\%} & 89.87ms & 612MB & \multirow{2}[1]{*}{1.13$\times$} & \multirow{2}[1]{*}{1.60\%} \\
          & Pruned Model & 150.97ms & 2034MB &       &       & 79.25ms & 602MB &       &  \\
    \bottomrule
    \end{tabular}}
  \label{tab:gpu time}%
  \vspace{-3mm}
\end{table}%

\subsection{Separation Effciency}
In \cref{sec:sota}, we have verified the effectiveness of \emph{SepPrune}. To further demonstrate the efficiency of \emph{SepPrune}, we measure the time and display memory overhead required for the pruned model during training and inference. Specifically, we perform the backward process (training) and forward process (inference) $1,000$ times on one second of audio at a sampling rate of $16$ kHz, and then take the average to represent the training and inference speeds. We report the GPU time and GPU display memory usage during training and inference, respectively. We utilize a single card when calculating GPU (NVIDIA A100) time. As shown in \cref{tab:gpu time}, \emph{SepPrune} not only brings effective training acceleration, but also significantly saves GPU memory during training (up to 50.2\%). Besides, \emph{SepPrune} achieves actual inference time acceleration, accelerating A-FRCNN-12 by $1.09$ times, TDANet by $1.08$ times, and SuDoRM-RF1.0x by $1.13$ times. Notably, the limited GPU display memory savings observed during inference stem from the fact that, in speech-separation models, the pruned parts themselves occupy only a small fraction of the total memory footprint. In summary, \emph{SepPrune} provides a practical solution for model acceleration.



\begin{table}[t]
  \centering
  \caption{Performance recovery after pruning with only $1$ epoch of fine-tuning on the LRS2-2Mix dataset. “Fine-tuning $1$ Epoch” denotes the pruned model with $1$ epoch fine-tuning. “Well-trained Model” denotes the pre-trained original model (without pruning).}
  \resizebox{0.99\textwidth}{!}{
    \begin{tabular}{c|cc|ccc|cc}
    \toprule
    \multirow{1.5}[4]{*}{Model} & \multicolumn{2}{c|}{Fine-tuning 1 Epoch} & \multicolumn{3}{c|}{Well-trained Model} & \multicolumn{2}{c}{Performance Recovery Rate} \\
\cmidrule{2-8}          & SDRi  & SI-SDRi & SDRi  & SI-SDRi & Training Epochs & SDRi  & SI-SDRi \\
    \midrule
    TDANet & 11.17 & 10.81 & 12.74 & 12.45 & 493   & 87.68\% & 86.83\% \\
    A-FRCNN-12 & 9.43  & 8.94  & 10.90  & 10.50  & 136   & 86.51\% & 85.14\% \\
    SuDoRM-RF1.0x & 5.18  & 4.06  & 11.43 & 11.10  & 86    & 45.32\% & 36.58\% \\
    \bottomrule
    \end{tabular}}
  \label{tab:1epoch}%
  \vspace{-3mm}
\end{table}%

\begin{table}[t]
  \centering
  \caption{Comparison of fine-tuning a pruned model and training a model of the same size from scratch. “Fine-tuning $1$ Epoch” denotes the pruned model with $1$ epoch fine-tuning. “Training 1 Epoch” means training a model of the same size as the pruned model from scratch for $1$ epochs. “Comparable Performance” denotes training a model from scratch with the same size as the pruned model achieves performance comparable to fine-tuning the pruned model for $1$ epoch.}
  \resizebox{0.99\textwidth}{!}{
    \begin{tabular}{c|cc|cc|ccc|c}
    \toprule
    \multirow{1.5}[4]{*}{Model} & \multicolumn{2}{c|}{Fine-tuning 1 Epoch} & \multicolumn{2}{c|}{Training 1 Epoch} & \multicolumn{3}{c|}{Comparable Performance} & \multirow{1.5}[4]{*}{Training Acceleration} \\
\cmidrule{2-8}          & SDRi  & SI-SDRi & SDRi  & SI-SDRi & SDRi  & SI-SDRi & Training Epochs &  \\
    \midrule
    TDANet & 11.17 & 10.81 & 4.31 & 2.80 &    11.13   &   10.75    &   36    &36$\times$  \\
    A-FRCNN-12 & 9.43  & 8.94  & 3.43  & 1.76  & 9.60 & 9.17 & 31    &  31$\times$\\
    SuDoRM-RF1.0x & 5.18  & 4.06  & 4.43 & 2.96 &  5.03     &  3.85     &  2     & 2$\times$ \\
    \bottomrule
    \end{tabular}}
  \label{tab:coverage}%
  \vspace{-5mm}
\end{table}%

\subsection{\emph{SepPrune} Enables Fast Convergence}
To further evaluate the efficiency of \emph{SepPrune} in actual pruning scenarios, we design two experiments to fine-tune pruned models with only $1$ epoch on the LRS2-2Mix dataset. These experiments are designed with the expectation that pruned models will recover most of their performance with minimal fine-tuning in real deployments. Specifically, we prune three typical speech separation models (TDANet, A-FRCNN-12 and SuDoRM-RF1.0x) and fine-tune them for $1$ epoch on the LRS2-2Mix dataset, and then compare them with the original unpruned models and retrained models (train from scratch) of the same size as the pruned models. 

As shown in \cref{tab:1epoch}, the original models take $493$ epochs (TDANet), $136$ epochs (A-FRCNN-12), and $86$ epochs (SuDoRM-RF1.0x) to complete training, while \emph{SepPrune} only fine-tune for $1$ epoch and restore the performance of most models to more than $85\%$, fully demonstrating the efficiency of \emph{SepPrune} in the training stage. Besides, we further explore whether it is more efficient to do a small amount of fine-tuning on the pruned model or to train a model of the same size from scratch under the same parameter budget. As shown in \cref{tab:coverage}, the performance of the randomly initialized small model after $1$ epoch of training is far behind the effect of fine-tuning the pruned model for $1$ epoch. Training a model of the same size as the pruned model directly from scratch to achieve the same effect as fine-tuning the pruned model for $1$ epochs requires dozens of epochs ($36$ for TDANet and $31$ for A-FRCNN-12), which fully demonstrates the huge efficiency advantage of \emph{SepPrune}. The performance recovery effect of SuDoRM-RF1.0x is significantly inferior to that of the other two models. We believe that this is mainly due to the fact that a large number of structures of the model are removed during the pruning process, making it difficult to quickly rebuild the model performance with only $1$ epoch of fine-tuning. Despite this, its performance is still better than a randomly initialized model of the same size trained from scratch for $1$ epoch, which shows that even if a large amount of structure is pruned, the retained pre-trained weights can still bring better initial performance than training from scratch with very limited epochs. In summary, \emph{SepPrune} can not only effectively restore model performance, but also significantly reduce costs by dozens of times of training acceleration.

\subsection{Ablation Study}
\label{sec:ablation}
We adopt the TDANet and A-FRCNN-12 trained on the LRS2-2Mix dataset in the ablation studies. The training configuration of ablation experiments is same as \cref{sec:Training and Evaluation}.

\textbf{Which optimization method is better: joint optimization of weights and masks, or step-by-step optimization?} To verify whether optimizing masks and weights step by step is better than joint optimization, we use the masks obtained by joint optimization and separate optimization for channel pruning respectively. As shown in \cref{tab:adastudy2}, the model obtained by step-by-step optimization achieves higher separation performance than the jointly‐optimized one, improving SDRi by $0.54$ dB and SI-SDRi by $0.72$ dB. We believe that this is because step-by-step optimization focuses on mask search first, making the preserved structure fit the task more accurately.

\begin{wraptable}{R}{7cm}
  \centering
  \vspace{-6mm}
  \caption{Importance of optimizing masks and weights in steps. Experiments are conducted using A-FRCNN-12 on the LRS2-2Mix dataset.}
    \begin{tabular}{c|cc}
    \toprule
    Method & SDRi  & SI-SDRi \\
    \midrule
    Joint Optimization & 12.05 & 11.53 \\
    Step-by-step Optimization & 12.59 & 12.25 \\
    \bottomrule
    \end{tabular}%
  \label{tab:adastudy2}%
\end{wraptable}

\textbf{The effect of mask learning with different numbers of iterations.} As we mentioned in ~\cref{sec:Training and Evaluation}, for cost-saving reasons, we only perform $500$ iterations of mask learning. Here, we conduct an ablation experiment with different iterations to evaluate the influence of the number of mask learning iterations on the final pruning effect. Specifically, we set the iteration to $\{300, 500,700,900,1100\}$ for experiments. As shown in \cref{tab:iteration}, different numbers of mask learning iterations do not affect the final performance or compression rate, so we set it to $500$ by default in this paper to minimize the training cost while ensuring the pruning effect.

\begin{wraptable}{R}{9cm}
  \centering
  \vspace{-6mm}
  \caption{Pruned models obtained by mask learning with different numbers of iterations. Experiments are conducted using TDANet on the LRS2-2Mix dataset.}
    \begin{tabular}{c|cccc}
    \toprule
    Iteration & Params & FLOPs & SDRi  & SI-SDRi \\
    \midrule
    300   & 1.91 M  & 4.32 GMac & 12.74 & 12.43 \\
    500   & 1.92 M & 4.33 GMac & 12.72 & 12.41 \\
    700   & 1.92 M & 4.33 GMac  & 12.68 & 12.40 \\
    900   & 1.92 M & 4.33 GMac  & 12.73 & 12.42 \\
    1100  &  1.90 M    &   4.31 GMac   &  12.71     & 12.41 \\
    \bottomrule
    \end{tabular}%
  \label{tab:iteration}%
\end{wraptable}

\textbf{The influence of the value of $\epsilon$.} As we mentioned in \cref{sec:method}, $\epsilon$ is a hyperparameter used to control the pruning ratio. Therefore, we set $\epsilon=\{0.5, 0.6,0.7,0.8,0.9\}$ to conduct experiments. As shown in \cref{tab:epsilon}, changing the value of $\epsilon$ can effectively adjust the model complexity and performance. In this study, when $\epsilon=0.7$, the model achieves a good balance between computational complexity and performance. Therefore, $\epsilon$ is set to $0.7$ by default.

\begin{wraptable}{R}{9cm}
  \centering
    \vspace{-6mm}
  \caption{Pruning models obtained with different $\epsilon$. Experiments are conducted using A-FRCNN-12 on the LRS2-2Mix dataset.}
    \begin{tabular}{ccccc}
    \toprule
    $\epsilon$ & Params & FLOPs & SDRi  & SI-SDRi \\
    \midrule
    0.5   & 3.69 M  & 20.25 GMac & 12.52 & 12.19 \\
    0.6   & 3.42 M  & 18.55 GMac & 12.58 & 12.23 \\
    0.7   & 3.06 M  & 16.52 GMac & 12.59 & 12.25 \\
    0.8   & 2.94 M  & 15.73 GMac & 12.35 & 12.03 \\
    0.9   & 2.71 M  & 14.49 GMac & 12.07 & 11.75 \\
    \bottomrule
    \end{tabular}%
  \label{tab:epsilon}%
\end{wraptable}


\section{Conclusion}
\label{sec:con}
In this paper, we have presented \emph{SepPrune}, the first pruning framework tailored specifically for deep speech separation models. \emph{SepPrune} first performs a structural calculation analysis on existing models to determine the layers with the highest computational cost. Subsequently, \emph{SepPrune} introduces differentiable masks to perform gradient-driven channel mask search and implements channel pruning based on the obtained masks. Experiments demonstrate that \emph{SepPrune} outperforms the existing channel pruning methods. Besides, \emph{SepPrune} can recover more than 85\% of the performance of the original model with just $1$ epoch of fine-tuning and converge much faster than training a model of the same size from scratch. Finally, \emph{SepPrune} offers a novel pruning paradigm for the design of lightweight speech separation models on devices with limited resources.

\textbf{Limitations.} Although this paper verifies the universality and effectiveness of \emph{SepPrune} on multiple mainstream models~\cite{li2022efficient,hu2021speech,tzinis2020sudo}, we have not yet conducted evaluations on the latest state-of-the-art models, such as Tiger~\cite{xu2024tiger} and SPMamba~\cite{li2024spmamba}. In the future, we will work on conducting experiments on more representative models to further verify the applicability and generalization ability of \emph{SepPrune}.

\clearpage
{
    \small
    \bibliographystyle{plain}
    \bibliography{ref}

\begin{thebibliography}{10}

\bibitem{achiam2023gpt}
Josh Achiam, Steven Adler, Sandhini Agarwal, Lama Ahmad, Ilge Akkaya, Florencia~Leoni Aleman, Diogo Almeida, Janko Altenschmidt, Sam Altman, Shyamal Anadkat, et~al.
\newblock Gpt-4 technical report.
\newblock {\em arXiv preprint arXiv:2303.08774}, 2023.

\bibitem{afouras2018deep}
Triantafyllos Afouras, Joon~Son Chung, Andrew Senior, Oriol Vinyals, and Andrew Zisserman.
\newblock Deep audio-visual speech recognition.
\newblock {\em IEEE transactions on pattern analysis and machine intelligence}, 44(12):8717--8727, 2018.

\bibitem{bai2022dual}
Yue Bai, Huan Wang, Zhiqiang Tao, Kunpeng Li, and Yun Fu.
\newblock Dual lottery ticket hypothesis.
\newblock {\em arXiv preprint arXiv:2203.04248}, 2022.

\bibitem{bengio2013estimating}
Yoshua Bengio, Nicholas L{\'e}onard, and Aaron Courville.
\newblock Estimating or propagating gradients through stochastic neurons for conditional computation.
\newblock {\em arXiv preprint arXiv:1308.3432}, 2013.

\bibitem{briere2008embedded}
Simon Briere, Jean-Marc Valin, Fran{\c{c}}ois Michaud, and Dominic L{\'e}tourneau.
\newblock Embedded auditory system for small mobile robots.
\newblock In {\em 2008 IEEE International Conference on Robotics and Automation}, pages 3463--3468. IEEE, 2008.

\bibitem{chang2017matterport3d}
Angel Chang, Angela Dai, Thomas Funkhouser, Maciej Halber, Matthias Niessner, Manolis Savva, Shuran Song, Andy Zeng, and Yinda Zhang.
\newblock Matterport3d: Learning from rgb-d data in indoor environments.
\newblock {\em arXiv preprint arXiv:1709.06158}, 2017.

\bibitem{chen2022soundspaces}
Changan Chen, Carl Schissler, Sanchit Garg, Philip Kobernik, Alexander Clegg, Paul Calamia, Dhruv Batra, Philip Robinson, and Kristen Grauman.
\newblock Soundspaces 2.0: A simulation platform for visual-acoustic learning.
\newblock {\em Advances in Neural Information Processing Systems}, 35:8896--8911, 2022.

\bibitem{chen2020dual}
Jingjing Chen, Qirong Mao, and Dong Liu.
\newblock Dual-path transformer network: Direct context-aware modeling for end-to-end monaural speech separation.
\newblock {\em arXiv preprint arXiv:2007.13975}, 2020.

\bibitem{chen2018shallowing}
Shi Chen and Qi~Zhao.
\newblock Shallowing deep networks: Layer-wise pruning based on feature representations.
\newblock {\em IEEE transactions on pattern analysis and machine intelligence}, 41(12):3048--3056, 2018.

\bibitem{chen2023rgp}
Zhuangzhi Chen, Jingyang Xiang, Yao Lu, Qi~Xuan, Zhen Wang, Guanrong Chen, and Xiaoniu Yang.
\newblock Rgp: Neural network pruning through regular graph with edges swapping.
\newblock {\em IEEE Transactions on Neural Networks and Learning Systems}, 2023.

\bibitem{cosentino2020librimix}
Joris Cosentino, Manuel Pariente, Samuele Cornell, Antoine Deleforge, and Emmanuel Vincent.
\newblock Librimix: An open-source dataset for generalizable speech separation.
\newblock {\em arXiv preprint arXiv:2005.11262}, 2020.

\bibitem{das2014you}
Anupam Das, Nikita Borisov, and Matthew Caesar.
\newblock Do you hear what i hear? fingerprinting smart devices through embedded acoustic components.
\newblock In {\em Proceedings of the 2014 ACM SIGSAC Conference on Computer and Communications Security}, pages 441--452, 2014.

\bibitem{fang2024tinyfusion}
Gongfan Fang, Kunjun Li, Xinyin Ma, and Xinchao Wang.
\newblock Tinyfusion: Diffusion transformers learned shallow.
\newblock {\em arXiv preprint arXiv:2412.01199}, 2024.

\bibitem{fang2024maskllm}
Gongfan Fang, Hongxu Yin, Saurav Muralidharan, Greg Heinrich, Jeff Pool, Jan Kautz, Pavlo Molchanov, and Xinchao Wang.
\newblock Maskllm: Learnable semi-structured sparsity for large language models.
\newblock {\em arXiv preprint arXiv:2409.17481}, 2024.

\bibitem{frankle2018lottery}
Jonathan Frankle and Michael Carbin.
\newblock The lottery ticket hypothesis: Finding sparse, trainable neural networks.
\newblock {\em arXiv preprint arXiv:1803.03635}, 2018.

\bibitem{gao2024bilevelpruning}
Shangqian Gao, Yanfu Zhang, Feihu Huang, and Heng Huang.
\newblock Bilevelpruning: unified dynamic and static channel pruning for convolutional neural networks.
\newblock In {\em Proceedings of the IEEE/CVF conference on computer vision and pattern recognition}, pages 16090--16100, 2024.

\bibitem{gumbel1954statistical}
Emil~Julius Gumbel.
\newblock {\em Statistical theory of extreme values and some practical applications: a series of lectures}, volume~33.
\newblock US Government Printing Office, 1954.

\bibitem{guo2020dmcp}
Shaopeng Guo, Yujie Wang, Quanquan Li, and Junjie Yan.
\newblock Dmcp: Differentiable markov channel pruning for neural networks.
\newblock In {\em Proceedings of the IEEE/CVF conference on computer vision and pattern recognition}, pages 1539--1547, 2020.

\bibitem{han2016eie}
Song Han, Xingyu Liu, Huizi Mao, Jing Pu, Ardavan Pedram, Mark~A Horowitz, and William~J Dally.
\newblock Eie: Efficient inference engine on compressed deep neural network.
\newblock {\em ACM SIGARCH Computer Architecture News}, 44(3):243--254, 2016.

\bibitem{he2016deep}
Kaiming He, Xiangyu Zhang, Shaoqing Ren, and Jian Sun.
\newblock Deep residual learning for image recognition.
\newblock In {\em Proceedings of the IEEE Conference on Computer Vision and Pattern Recognition}, pages 770--778, 2016.

\bibitem{he2017channel}
Yihui He, Xiangyu Zhang, and Jian Sun.
\newblock Channel pruning for accelerating very deep neural networks.
\newblock In {\em Proceedings of the IEEE international conference on computer vision}, pages 1389--1397, 2017.

\bibitem{hoang2023revisiting}
Duc~NM Hoang and Shiwei Liu.
\newblock Revisiting pruning at initialization through the lens of ramanujan graph.
\newblock {\em ICLR 2023}, 2023.

\bibitem{hu2021speech}
Xiaolin Hu, Kai Li, Weiyi Zhang, Yi~Luo, Jean-Marie Lemercier, and Timo Gerkmann.
\newblock Speech separation using an asynchronous fully recurrent convolutional neural network.
\newblock {\em Advances in Neural Information Processing Systems}, 34:22509--22522, 2021.

\bibitem{jang2016categorical}
Eric Jang, Shixiang Gu, and Ben Poole.
\newblock Categorical reparameterization with gumbel-softmax.
\newblock {\em arXiv preprint arXiv:1611.01144}, 2016.

\bibitem{jiang2025dual}
Xilin Jiang, Cong Han, and Nima Mesgarani.
\newblock Dual-path mamba: Short and long-term bidirectional selective structured state space models for speech separation.
\newblock In {\em ICASSP 2025-2025 IEEE International Conference on Acoustics, Speech and Signal Processing (ICASSP)}, pages 1--5. IEEE, 2025.

\bibitem{kingma2014adam}
Diederik~P Kingma and Jimmy Ba.
\newblock Adam: A method for stochastic optimization.
\newblock {\em arXiv preprint arXiv:1412.6980}, 2014.

\bibitem{le2019sdr}
Jonathan Le~Roux, Scott Wisdom, Hakan Erdogan, and John~R Hershey.
\newblock Sdr--half-baked or well done?
\newblock In {\em ICASSP 2019-2019 IEEE International Conference on Acoustics, Speech and Signal Processing (ICASSP)}, pages 626--630. IEEE, 2019.

\bibitem{li2016pruning}
Hao Li, Asim Kadav, Igor Durdanovic, Hanan Samet, and Hans~Peter Graf.
\newblock Pruning filters for efficient convnets.
\newblock {\em arXiv preprint arXiv:1608.08710}, 2016.

\bibitem{li2024spmamba}
Kai Li, Guo Chen, Runxuan Yang, and Xiaolin Hu.
\newblock Spmamba: State-space model is all you need in speech separation.
\newblock {\em arXiv preprint arXiv:2404.02063}, 2024.

\bibitem{li2024subnetwork}
Kai Li and Yi~Luo.
\newblock Subnetwork-to-go: Elastic neural network with dynamic training and customizable inference.
\newblock In {\em ICASSP 2024-2024 IEEE International Conference on Acoustics, Speech and Signal Processing (ICASSP)}, pages 6775--6779. IEEE, 2024.

\bibitem{li2022efficient}
Kai Li, Runxuan Yang, and Xiaolin Hu.
\newblock An efficient encoder-decoder architecture with top-down attention for speech separation.
\newblock {\em arXiv preprint arXiv:2209.15200}, 2022.

\bibitem{li2024sglp}
Yuqi Li, Yao Lu, Zeyu Dong, Chuanguang Yang, Yihao Chen, and Jianping Gou.
\newblock Sglp: A similarity guided fast layer partition pruning for compressing large deep models.
\newblock {\em arXiv preprint arXiv:2410.14720}, 2024.

\bibitem{lin2020hrank}
Mingbao Lin, Rongrong Ji, Yan Wang, Yichen Zhang, Baochang Zhang, Yonghong Tian, and Ling Shao.
\newblock Hrank: Filter pruning using high-rank feature map.
\newblock In {\em Proceedings of the IEEE/CVF conference on computer vision and pattern recognition}, pages 1529--1538, 2020.

\bibitem{lin2020channel}
Mingbao Lin, Rongrong Ji, Yuxin Zhang, Baochang Zhang, Yongjian Wu, and Yonghong Tian.
\newblock Channel pruning via automatic structure search.
\newblock {\em arXiv preprint arXiv:2001.08565}, 2020.

\bibitem{ling2024slimgpt}
Gui Ling, Ziyang Wang, and Qingwen Liu.
\newblock Slimgpt: Layer-wise structured pruning for large language models.
\newblock {\em Advances in Neural Information Processing Systems}, 37:107112--107137, 2024.

\bibitem{liu2022unreasonable}
Shiwei Liu, Tianlong Chen, Xiaohan Chen, Li~Shen, Decebal~Constantin Mocanu, Zhangyang Wang, and Mykola Pechenizkiy.
\newblock The unreasonable effectiveness of random pruning: Return of the most naive baseline for sparse training.
\newblock {\em arXiv preprint arXiv:2202.02643}, 2022.

\bibitem{liu2021swin}
Ze~Liu, Yutong Lin, Yue Cao, Han Hu, Yixuan Wei, Zheng Zhang, Stephen Lin, and Baining Guo.
\newblock Swin transformer: Hierarchical vision transformer using shifted windows.
\newblock In {\em Proceedings of the IEEE/CVF international conference on computer vision}, pages 10012--10022, 2021.

\bibitem{liu2019metapruning}
Zechun Liu, Haoyuan Mu, Xiangyu Zhang, Zichao Guo, Xin Yang, Kwang-Ting Cheng, and Jian Sun.
\newblock Metapruning: Meta learning for automatic neural network channel pruning.
\newblock In {\em Proceedings of the IEEE/CVF international conference on computer vision}, pages 3296--3305, 2019.

\bibitem{lu2024reassessing}
Yao Lu, Hao Cheng, Yujie Fang, Zeyu Wang, Jiaheng Wei, Dongwei Xu, Qi~Xuan, Xiaoniu Yang, and Zhaowei Zhu.
\newblock Reassessing layer pruning in llms: New insights and methods.
\newblock {\em arXiv preprint arXiv:2411.15558}, 2024.

\bibitem{lu2022understanding}
Yao Lu, Wen Yang, Yunzhe Zhang, Zuohui Chen, Jinyin Chen, Qi~Xuan, Zhen Wang, and Xiaoniu Yang.
\newblock Understanding the dynamics of dnns using graph modularity.
\newblock In {\em European Conference on Computer Vision}, pages 225--242. Springer, 2022.

\bibitem{lu2024generic}
Yao Lu, Yutao Zhu, Yuqi Li, Dongwei Xu, Yun Lin, Qi~Xuan, and Xiaoniu Yang.
\newblock A generic layer pruning method for signal modulation recognition deep learning models.
\newblock {\em IEEE Transactions on Cognitive Communications and Networking}, 2024.

\bibitem{luo2019conv}
Yi~Luo and Nima Mesgarani.
\newblock Conv-tasnet: Surpassing ideal time--frequency magnitude masking for speech separation.
\newblock {\em IEEE/ACM transactions on audio, speech, and language processing}, 27(8):1256--1266, 2019.

\bibitem{luo2023music}
Yi~Luo and Jianwei Yu.
\newblock Music source separation with band-split rnn.
\newblock {\em IEEE/ACM Transactions on Audio, Speech, and Language Processing}, 31:1893--1901, 2023.

\bibitem{ma2023llm}
Xinyin Ma, Gongfan Fang, and Xinchao Wang.
\newblock Llm-pruner: On the structural pruning of large language models.
\newblock {\em Advances in neural information processing systems}, 36:21702--21720, 2023.

\bibitem{maryn2017mobile}
Youri Maryn, Femke Ysenbaert, Andrzej Zarowski, and Robby Vanspauwen.
\newblock Mobile communication devices, ambient noise, and acoustic voice measures.
\newblock {\em Journal of Voice}, 31(2):248--e11, 2017.

\bibitem{panayotov2015librispeech}
Vassil Panayotov, Guoguo Chen, Daniel Povey, and Sanjeev Khudanpur.
\newblock Librispeech: an asr corpus based on public domain audio books.
\newblock In {\em 2015 IEEE international conference on acoustics, speech and signal processing (ICASSP)}, pages 5206--5210. IEEE, 2015.

\bibitem{park2016faster}
Jongsoo Park, Sheng Li, Wei Wen, Ping Tak~Peter Tang, Hai Li, Yiran Chen, and Pradeep Dubey.
\newblock Faster cnns with direct sparse convolutions and guided pruning.
\newblock {\em arXiv preprint arXiv:1608.01409}, 2016.

\bibitem{qian2024reasoning}
Rui Qian, Xin Yin, and Dejing Dou.
\newblock Reasoning to attend: Try to understand how< seg> token works.
\newblock {\em arXiv preprint arXiv:2412.17741}, 2024.

\bibitem{series2011algorithms}
BS~Series.
\newblock Algorithms to measure audio programme loudness and true-peak audio level.
\newblock {\em International Telecommunication Union Radiocommunication Assembly}, 2011.

\bibitem{simonyan2014deep}
Karen Simonyan and Andrew Zisserman.
\newblock Very deep convolutional networks for large-scale image recognition.
\newblock {\em arXiv preprint arXiv:1409.1556}, 2014.

\bibitem{subakan2021attention}
Cem Subakan, Mirco Ravanelli, Samuele Cornell, Mirko Bronzi, and Jianyuan Zhong.
\newblock Attention is all you need in speech separation.
\newblock In {\em ICASSP 2021-2021 IEEE International Conference on Acoustics, Speech and Signal Processing (ICASSP)}, pages 21--25. IEEE, 2021.

\bibitem{subakan2022real}
Cem Subakan, Mirco Ravanelli, Samuele Cornell, and Fran{\c{c}}ois Grondin.
\newblock Real-m: Towards speech separation on real mixtures.
\newblock In {\em ICASSP 2022-2022 IEEE International Conference on Acoustics, Speech and Signal Processing (ICASSP)}, pages 6862--6866. IEEE, 2022.

\bibitem{sun2023simple}
Mingjie Sun, Zhuang Liu, Anna Bair, and J~Zico Kolter.
\newblock A simple and effective pruning approach for large language models.
\newblock {\em arXiv preprint arXiv:2306.11695}, 2023.

\bibitem{tang2023sr}
Hui Tang, Yao Lu, and Qi~Xuan.
\newblock Sr-init: An interpretable layer pruning method.
\newblock In {\em ICASSP 2023-2023 IEEE International Conference on Acoustics, Speech and Signal Processing (ICASSP)}, pages 1--5. IEEE, 2023.

\bibitem{touvron2023llama}
Hugo Touvron, Thibaut Lavril, Gautier Izacard, Xavier Martinet, Marie-Anne Lachaux, Timoth{\'e}e Lacroix, Baptiste Rozi{\`e}re, Naman Goyal, Eric Hambro, Faisal Azhar, et~al.
\newblock Llama: Open and efficient foundation language models.
\newblock {\em arXiv preprint arXiv:2302.13971}, 2023.

\bibitem{tzinis2020sudo}
Efthymios Tzinis, Zhepei Wang, and Paris Smaragdis.
\newblock Sudo rm-rf: Efficient networks for universal audio source separation.
\newblock In {\em 2020 IEEE 30th International Workshop on Machine Learning for Signal Processing (MLSP)}, pages 1--6. IEEE, 2020.

\bibitem{vincent2006performance}
Emmanuel Vincent, R{\'e}mi Gribonval, and C{\'e}dric F{\'e}votte.
\newblock Performance measurement in blind audio source separation.
\newblock {\em IEEE transactions on audio, speech, and language processing}, 14(4):1462--1469, 2006.

\bibitem{wang2024rl}
Boyao Wang and Volodymyr Kindratenko.
\newblock Rl-pruner: Structured pruning using reinforcement learning for cnn compression and acceleration.
\newblock {\em arXiv preprint arXiv:2411.06463}, 2024.

\bibitem{wang2021recent}
Huan Wang, Can Qin, Yue Bai, Yulun Zhang, and Yun Fu.
\newblock Recent advances on neural network pruning at initialization.
\newblock {\em arXiv preprint arXiv:2103.06460}, 2021.

\bibitem{wang2023ntk}
Yite Wang, Dawei Li, and Ruoyu Sun.
\newblock Ntk-sap: Improving neural network pruning by aligning training dynamics.
\newblock {\em arXiv preprint arXiv:2304.02840}, 2023.

\bibitem{wang2023tf}
Zhong-Qiu Wang, Samuele Cornell, Shukjae Choi, Younglo Lee, Byeong-Yeol Kim, and Shinji Watanabe.
\newblock Tf-gridnet: Making time-frequency domain models great again for monaural speaker separation.
\newblock In {\em ICASSP 2023-2023 IEEE international conference on acoustics, speech and signal processing (ICASSP)}, pages 1--5. IEEE, 2023.

\bibitem{wichern2019wham}
Gordon Wichern, Joe Antognini, Michael Flynn, Licheng~Richard Zhu, Emmett McQuinn, Dwight Crow, Ethan Manilow, and Jonathan~Le Roux.
\newblock Wham!: Extending speech separation to noisy environments.
\newblock {\em arXiv preprint arXiv:1907.01160}, 2019.

\bibitem{wu2023efficient}
Jie Wu, Dingshun Zhu, Leyuan Fang, Yue Deng, and Zhun Zhong.
\newblock Efficient layer compression without pruning.
\newblock {\em IEEE Transactions on Image Processing}, 32:4689--4700, 2023.

\bibitem{xu2024tiger}
Mohan Xu, Kai Li, Guo Chen, and Xiaolin Hu.
\newblock Tiger: Time-frequency interleaved gain extraction and reconstruction for efficient speech separation.
\newblock {\em arXiv preprint arXiv:2410.01469}, 2024.

\bibitem{yang2022tfpsnet}
Lei Yang, Wei Liu, and Weiqin Wang.
\newblock Tfpsnet: Time-frequency domain path scanning network for speech separation.
\newblock In {\em ICASSP 2022-2022 IEEE International Conference on Acoustics, Speech and Signal Processing (ICASSP)}, pages 6842--6846. IEEE, 2022.

\bibitem{yang2025wanda++}
Yifan Yang, Kai Zhen, Bhavana Ganesh, Aram Galstyan, Goeric Huybrechts, Markus M{\"u}ller, Jonas~M K{\"u}bler, Rupak~Vignesh Swaminathan, Athanasios Mouchtaris, Sravan~Babu Bodapati, et~al.
\newblock Wanda++: Pruning large language models via regional gradients.
\newblock {\em arXiv preprint arXiv:2503.04992}, 2025.

\bibitem{zeghidour2021wavesplit}
Neil Zeghidour and David Grangier.
\newblock Wavesplit: End-to-end speech separation by speaker clustering.
\newblock {\em IEEE/ACM Transactions on Audio, Speech, and Language Processing}, 29:2840--2849, 2021.

\bibitem{zhang2021transmask}
Zining Zhang, Bingsheng He, and Zhenjie Zhang.
\newblock Transmask: A compact and fast speech separation model based on transformer.
\newblock In {\em ICASSP 2021-2021 IEEE International Conference on Acoustics, Speech and Signal Processing (ICASSP)}, pages 5764--5768. IEEE, 2021.

\bibitem{zhuang2018discrimination}
Zhuangwei Zhuang, Mingkui Tan, Bohan Zhuang, Jing Liu, Yong Guo, Qingyao Wu, Junzhou Huang, and Jinhui Zhu.
\newblock Discrimination-aware channel pruning for deep neural networks.
\newblock {\em Advances in neural information processing systems}, 31, 2018.

\end{thebibliography}
}

\end{document}